\documentclass[prl,twocolumn,showpacs,nobibnotes,floatfix,superscriptaddress]{revtex4}

\usepackage{graphicx,eucal,amsfonts}

\begin{document}

\title{Matrix Element Distributions as a Signature of Entanglement Generation}

\author{Yaakov S. Weinstein}
\thanks{To whom correspondence should be addressed. \\ Present address: Quantum Information Science Group, MITRE,
Eatontown, NJ 07724}
\email{weinstein@mitre.org}
\author{C. Stephen Hellberg}
\email{hellberg@dave.nrl.navy.mil}
\affiliation{Center for Computational Materials Science, Naval Research Laboratory, Washington, DC 20375}

\begin{abstract}
We explore connections between an operator's matrix element distribution and
its entanglement generation. Operators with matrix element distributions 
similar to those of random matrices generate states of high multi-partite 
entanglement. This occurs even when other statistical properties of the 
operators do not conincide with random matrices. Similarly, operators with 
some statistical properties of random matrices may not exhibit random matrix 
element distributions and will not produce states with high levels of 
multi-partite entanglement. Finally, we show that operators with similar 
matrix element distributions generate similar amounts of entanglement.
\end{abstract}

\pacs{03.67.Mn  
      05.45.Mt  
      03.67.Lx} 
\maketitle

Entanglement, correlations between quantum systems beyond what is 
classically possible, is an essential phenomenon in quantum information 
processing and a necessary resource for quantum communication. In the space
of pure states, the overwhelming majority are of high multi-partite 
entanglement with respect to a qubit architecture \cite{Scott}. Such states 
are necessary for quantum protocols calling for random, highly entangled states
including superdense coding \cite{Aram}, remote state preparation 
\cite{Bennet}, and data hiding schemes \cite{Hayden}. 

Random states can be produced on a quantum computer by applying random unitary
operators to computational basis states. However, the implementation of 
operators drawn randomly from the space of all unitary operators, the circular 
unitary ensemble (CUE), is very inefficient. Instead, other operators have 
been suggested as possibly efficient substitutes for the production of 
random, highly entangled states. These include quantum chaotic 
operators \cite{WGSH} and pseudo-random operators \cite{RM,QCARM,CRM}. 
However, these operators are not truly random and thus do not 
uniformly cover the space of pure states. Here we attempt to identify what 
statistical properties of random matrices lead to the production of highly 
entangled, random states. Identification of such a link between entanglement 
and randomness provides a deeper understanding of entanglement. 
In addition, isolating these properties may focus the search for 
operators with the ability to efficiently produce random states. Such
operators need only fulfill the identified statistical properties and
can fall short of CUE in regards to other statistical distributions.

In this paper we suggest that the element distribution of a given operator
is a vital statistical property when attempting to create highly entangled 
states. To demonstrate this we first show that the larger percentage of CUE
covered by a given ensemble, the more the matrix element distribution 
converges to CUE and so does the entanglement generation. Of course, 
this will generally cause other properties to also approach CUE. We then
isolate various statistical properties through the use of operators which 
have certain statistical properties similar to CUE but not others. This 
disconnects the matrix elements from other statistical properties and 
shows their primacy in entanglement production. Specifically, an operator may 
not produce CUE-levels of entanglement if the matrix element distribution 
does not follow CUE, despite having other statistical similarities to CUE. 
Also, CUE-like entanglement production can be achieved with operators that 
have CUE-like matrix element distributions even if other statistical 
properties do not follow CUE. We note that not all of the operators we 
explore can be efficiently implemented on a quantum computer. Rather, the 
goal is to understand how certain operators can produce states with high 
levels of entanglement. This work concentrates on the matrix element 
distribution without exploring higher order correlations between the elements. 
These correlation may also play an important role in entanglement generation 
and will be the subject of further study. We have briefly mentioned the 
importance of the matrix elements to entanglement generation in Ref. \cite{YSW}.

The operator classes used in this work to demonstrate all of the above, 
are (1) the interpolating ensembles \cite{Zyc3}, a one-parameter family of 
ensembles which interpolate between diagonal matrices with uniform, 
independently distributed elements, and CUE, (2) pseudo-random operators 
\cite{RM} proposed as possibly efficient substitutes for random matrices, 
and (3) quantum chaotic operators which are generally known to have many 
statistical properties similar to random matrices \cite{BGS}.

CUE matrices can be generated by multiplying eigenvectors of a Hermitian 
matrix belonging to the Gaussian unitary ensemble (GUE) by a random phase and 
using the resulting vectors as the CUE matrix columns \cite{Zyc2}. 
Thus, the squared modulus, or amplitude, of CUE matrix elements follows a 
distribution equal to that of GUE eigenvector element amplitudes. Let 
$c^l_k$ denote the $k$th component of the $l$th GUE eigenvector. 
The distribution of amplitudes, $\eta = |c^l_k|^2$, is 
\begin{equation}
\tilde{P}_{GUE}(\eta) = (N-1)(1-\eta)^{N-2},
\end{equation}
where $N$ is the Hilbert space dimension. In the limit $N \rightarrow \infty$, 
after rescaling to unit mean, the distribution is given by 
\begin{equation}
P_{GUE}(y) = e^{-y}, 
\end{equation}
where $y = N\eta$ \cite{Zyc}. Since $\eta$ is unchanged when multiplied 
by a phase, the distribution, $P_{CUE}(x)$, of the rescaled amplitude of 
CUE matrix elements $x$, is equal to $P_{GUE}(y)$.

As a practical measure of multi-partite entanglement for an $n$-qubit system, 
we explore the average bipartite entanglement between each qubit and the rest 
of the system 
\cite{Meyer,Bren2},
\begin{equation}
\label{Q}
Q = 2-\frac{2}{n}\sum^n_{j=1}Tr[\rho_j^2],
\end{equation}
where $\rho_j$ is the reduced density matrix of qubit $j$. 
In this work, we study the distribution of $Q$ after one iteration of an
operator as compared to the distribution of $Q$ for CUE matrices, $P_{CUE}(Q)$,
and the average entanglement as a function of time, $\langle Q(t) \rangle$, 
compared to the CUE average entanglement \cite{Scott} 
\begin{equation}
\langle Q \rangle_{CUE}= (N-2)/(N+1).
\end{equation}

CUE matrices can also be generated based on the Hurwitz parameterization 
\cite{Zyc2} and a modification of this construction is used to generate the 
interpolating ensembles. The interpolating ensembles are a one-parameter,  
family of ensembles which interpolate between diagonal matrices with uniform, 
independently distributed elements, and CUE. The parameter, $\delta$, can take
on any value between 0, representing diagonal matrices, and 1, for CUE 
matrices. The exact method for the construction of CUE matrices via the Hurwitz
parameterization and the modifications needed for the interpolating ensembles 
is reviewed in Appendix A. Interpolating ensembles have proved useful in 
analyzing certain scattering matrices \cite{Zyc4,SB} and a quantum 
electron pump \cite{CB}. One of the attractive features of these ensembles 
is that their eigenvalue and eigenvector properties have only a weak 
dependence on $N$. The matrix elements for these operators, however, show a 
strong dependence on $N$ and this, in turn, effects their entangling power.

Pseudo-random matrices \cite{RM,CRM,QCARM} are operators that were proposed 
as possible efficient replacements of inefficient random operators in quantum 
information protocols. To implement a pseudo-random operator apply $m$ 
iterations of the $n$ qubit gate: random $SU(2)$ rotation to each qubit, 
then evolve the system via all nearest neighbor couplings \cite{RM}. A random 
$SU(2)$ rotation is described by Eqs. (\ref{E1}) and (\ref{E3}). The nearest 
neighbor coupling operator used is:
\begin{equation}
U_{nnc} = \exp(i\frac{\pi}{4}\sum^{n-1}_{j=1}\sigma_z^j\otimes\sigma_z^{j+1}),
\end{equation}
where $\sigma_z^j$ is the $j$th qubit $z$-direction Pauli spin operator.
The random rotations are different for each qubit and each iteration, but the
coupling constant is always $\pi/4$ to maximize entanglement generation. After 
the $m$ iterations, a final set of random rotations is applied. 

\begin{figure}
\includegraphics[height=5.8cm, width=8cm]{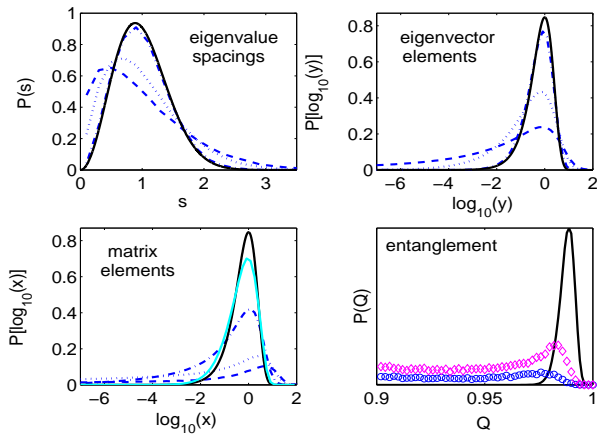}
\caption{\label{CUEdelta}
(Color online) Distributions of nearest neighbor eigenvalue spacings 
(upper-left) and eigenvector element amplitudes (upper-right) for matrices 
of the interpolating ensembles with $\delta = .1$ (dashed), .5 (dotted) 
and .9 (chained). The $N = 256$ matrix element plot (lower-left) 
includes the distribution for $\delta = .98$ (light solid line). For this 
$\delta$ the eigenvalue and eigenvector distributions are indistinguishable 
 from random (solid line) for the resolution of the figure. The matrix element 
distribution appears to converge more slowly which may be manifest in the 
entanglement generated by operating with 100 8-qubit $\delta = .9$ ($\bigcirc$), 
and $.98$ ($\diamond$) matrices on all computational basis states (lower right). 
}
\end{figure}

\begin{figure}
\includegraphics[height=5.8cm, width=8cm]{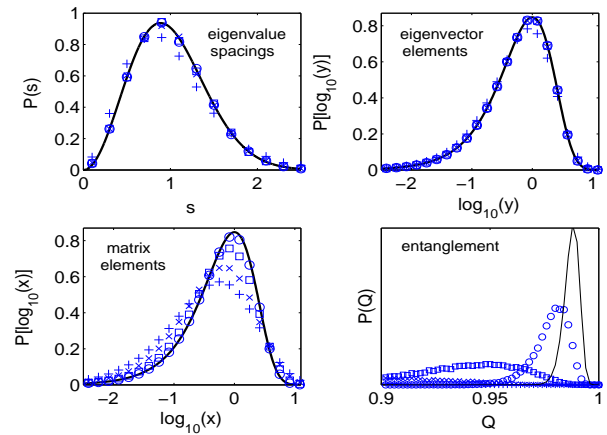}
\caption{\label{PR}
(Color online) Distribution of nearest neighbor eigenvalue spacings 
(top left), eigenvector elements (top right), matrix elements (bottom left), 
and $Q$ (bottom right) for $N = 256$ pseudo-random maps of $m = 2$ (+), 
4 ($\times$), 8 ($\square$), and 16 ($\bigcirc$). The eigenvalue and 
eigenvector distributions appear to converge to that of CUE (solid lines) 
more quickly than the matrix element distribution and entanglement 
distribution. To approach $P_{CUE}(Q)$ with one iteration of a map requires 
$m \simeq 40$ \protect\cite{RM,YSW}.
}
\end{figure}

Reference \cite{YSW} discusses various statistical properties of both 
these operator classes in connection with entanglement production. Some
of these properties are displayed here for completion. Figs. \ref{CUEdelta} 
and \ref{PR} show nearest neighbor eigenangle spacings (the more intricate 
number variance is provided in Ref. \cite{YSW}), eigenvalue element 
distribution, matrix element distribution and one-iteration entanglement 
distribution for the interpolating ensemble matrices and pseudo-random 
operators for constant $N$. As the operators cover more of CUE, 
$\delta \rightarrow 1$ and $m \rightarrow \infty$, the various properties 
including the matrix element distribution and the entanglement generation 
approach CUE distributions, as expected. We note however, that the matrix 
elements and entanglement generation appear to approach CUE more slowly than 
the other properties leading us to suspect that there may be a connection. 
This is born out by rewriting the average of $Q$ in terms of the elements of 
the wavefunction 
\begin{equation}
\langle Q \rangle = 4\big(\sum^{N/2}_{m = 1}\sum^N_{n = \frac{N}{2}+1}\langle|c_m|^2|c_n|^2\rangle - 
\sum^{N/2}_{q = 1}\langle |c_q|^2|c_{q+\frac{N}{2}}|^2\rangle\big),
\label{Qavg}
\end{equation}
where $c_i$ are the elements of the wavefunction. When applying an operator
to an intitial computational basis states, the output wavefunction elements are 
equivalent to the operator elements. To properly assert the primacy of this 
connection between matrix elements and entanglement generation we investigate 
groups of operators that fulfill only some statistical properties of CUE, but 
not others, and in that way isolate the property that causes CUE-like entanglement 
generation.

First, we show that the eigenvalue spectrum alone cannot be the sole cause of 
entanglement generation. This is done in two ways: by identifying a 
set of operators that have eigenvalues with statistical properties that 
match CUE but generate no entanglement, and, second, by identifying operators that 
do not have CUE eigenvalue properties but nevertheless generate CUE levels 
of entanglement. The first set is that of diagonal operators in which the 
elements are the eigenvalues of a CUE operator. When applying diagonal 
operators to computational basis states no entanglement is generated. 
The eigenvector element distribution and matrix element distribution for 
these diagonal operators clearly do not follow the CUE distributions. 
Nevertheless, the eigenvalue spectra fulfill all statistical properties of 
CUE including nearest-neighbor spacings and higher order correlation 
functions. The second set of operators are created as follows: let $D$ be a 
unitary diagonal operator with random phases drawn uniformly from 0 to 
$2\pi$. Operators $U = U_{CUE}DU_{CUE}^{\dag}$ have eigenvector distributions 
that follow CUE, matrix element distributions that follow CUE, and 
entanglement generation equal to CUE. Yet, the nearest-neighbor eigenvalue 
distribution follows a Poissonian and not the Wigner-Dyson 
distribution \cite{Wigner,Dyson}. What we see from these types of 
operators is that the eigenvalue distribution of an operator is not a 
primary factor in an operators' entanglement generation. This is not to say 
that operators with high entanglement generation never follow the 
Wigner-Dyson distribution, as we will see they usually do. Rather, the 
eigenvalue distribution is not the defining property generating entanglement.

The next step is to divorce the matrix element distribution and entanglement
generation of an operator from its eigenvector distribution. This 
cannot be done completely. Ref. \cite{DK} proves that the entanglement of the 
eigenvectors generally provides a lower bound for the asymptotic time value of 
entanglement generation. This proof is done for bipartite entanglement and 
can be extended to our investigations because $Q$ is merely the average of 
bipartite entanglement between each individual qubit and the rest of the 
system. This bound, however, does not exclude the possibility of 
a high entanglement generation for operators without random eigenvectors. 
Nor does this result tell how long it takes to reach this asymptotic value, 
a necessary question when trying to create a highly entangled state on a 
quantum computer. Through the use of the interpolating ensembles discussed 
above we show how the matrix element distribution of an operator relates to 
these issues. 

\begin{figure}
\includegraphics[height=5.8cm, width=8cm]{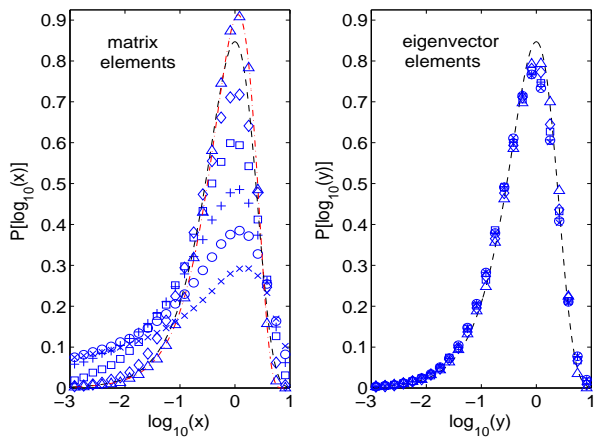}
\caption{\label{N}
(Color online) Matrix element distribution (left) and eigenvector element 
distribution (right) for interpolating ensemble matrices with $\delta = .9$
and $N = 256$ ($\times$), 128 ($\bigcirc$), 64 ($+$), 32 ($\square$), 16 
($\diamond$), and 8 ($\triangle$). The matrix element distributions gets
further and further from that expected of CUE (dashed line for the 
$N \rightarrow \infty$ limit and chained line for $N = 8$) as $N$ increases. 
This is in contrast with the eigenvector element distribution in which is 
remarkably stable as a function of $N$.
}
\end{figure}

As mentioned, many of the statistical properties of interpolating ensemble 
matrices are only weakly dependent on $N$. However, the matrix element 
distribution and the entanglement generation are strongly dependent on $N$. 
Fig. \ref{N} shows the eigenvector element distributions and matrix element 
distributions for $\delta = .9$ interpolating operators at various values 
of $N$. As $N$ decreases the matrix element distributions approach the CUE 
distribution. The eigenvector elements on the other hand, are practically 
constant with $N$.

\begin{figure}
\includegraphics[height=5.8cm, width=8cm]{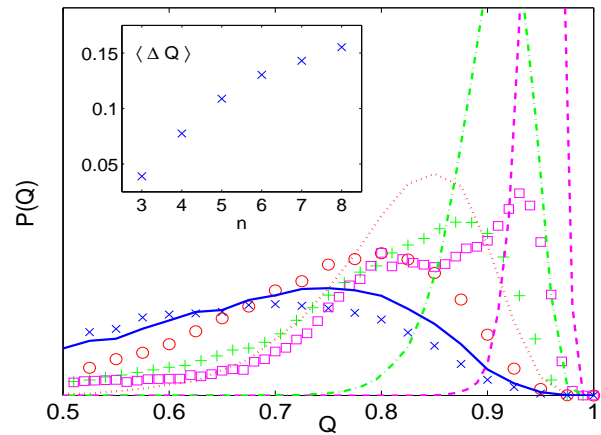}
\caption{\label{QN} 
(Color online) Entanglement spectra for CUE operators (lines) and 
$\delta = .9$ interpolating ensemble operators (shapes), for $N = 8$ 
(solid line and $\times$), 16 (dotted line and $\bigcirc$), 32 (chained line 
and +) and 64 (dashed line and $\square$). As $N$ increases the distributions 
of the interpolating ensemble operators diverges further from the CUE 
distribution. The inset shows the difference between average entanglement 
generation for CUE operators and the $\delta = .9$ interpolating ensemble 
operators, $\Delta Q$, as a function of number of qubits, $n = log_2(N)$. 
As $N$ increases the average entanglement production 
gets further from the CUE average. 
}
\end{figure}

The average entanglement generated in one iteration of the operator, 
compared to that expected from CUE matrices, is shown in Fig. \ref{QN} as 
a function of $N$ for the same $\delta = .9$ operators. As $N$ gets smaller 
the average entanglement generated approaches that of CUE, in a way similar 
to the matrix element distribution. This despite the fact that the eigenvector 
element distribution remains constant as a function of $N$. For large $\delta$ 
interpolating ensemble, say $\delta = .99$, the eigenvector and eigenvalue 
distributions will be practically indistinguishable 
from CUE. However, as we see here, the matrix element distribution, and thus 
the entanglement generation, will fall short of CUE. The larger the Hilbert
space the further from CUE. The interpolating ensemble operators thus provide 
some divergence between the eigenvector distribution and the matrix element 
distribution. For small $N$ the matrix element distribution and entanglement 
generation are practically random though the eigenvector element distribution 
is not. For large $N$ and large $\delta$ the eigenvector element distribution is 
practically random but the matrix element distribution and entanglement generation
are not. This further demonstrates the importance of an operators' matrix elements 
in entanglement generation. The time evolution of these operators is explored 
below.

A quantum computer programmer starting with a computational basis state and 
attempting to generate a random state of high multi-partite entanglement will 
want to apply an operator with a matrix element distribution as close as 
possible to CUE. Most likely the operator will also exhibit other statistical 
properties close to CUE, as occurs with pseudo-random operators 
\cite{YSW}, but that need not be the case. 

Identifying one statistical property that is the dominant cause of high 
entanglement generation is also important for understanding entanglement 
as a quantum phenomenon and its relation to quantum chaos.
Quantum chaotic operators are known to exhibit many properties of random 
matrices \cite{BGS,Haake} including the ability to produce entanglement. 
Numerical simulations of two coupled subsystems demonstrate the greater 
entanglement generation of chaotic versus regular quantum dynamics 
\cite{L1,FNP,MS,WGSH} and analytical results have been obtained through 
various methods \cite{TFM,L2,J}. As with the previous classes of operators, 
we are interested in a quantum chaotic operator's ability to produce highly 
entangled states from initial computational basis states. Applying a chaotic 
operator once will not, in general, produce entanglement on par with random 
operators \cite{Scott,WGSH}. This is in line with the deviant short time 
behavior of chaotic systems with respect to other statistical properties 
such as the level or number variance. The deviant behavior is attributed 
to short periodic phase space orbits \cite{AS}. Long time entanglement 
generation behavior is related to the operator's eigenvectors \cite{DK}. 
We have already demonstrated that the operator eigenvalues and eigenvectors 
are not primary with respect to entanglement generation, thus we demonstrate 
the observed short and long time entanglement generation behavior is also 
reflected in the matrix element distribution.

Upon increasing the number of iterations, $t$, of a quantum chaotic operator
the average entanglement of initial computational basis states, 
$\langle Q(t)\rangle$, can approach that of random operators. Similarly, the 
matrix element distribution of chaotic operators at higher powers approaches the 
CUE distribution. To demonstrate this we revisit the entanglement production 
of the \cite{BV} quantum baker's map \cite{Scott} and explore other quantized 
chaotic maps. 

\begin{figure}
\includegraphics[height=5.8cm, width=8cm]{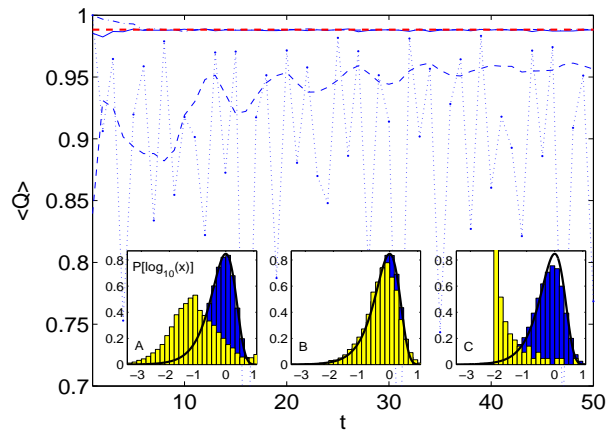}
\caption{\label{M4}
(Color online) Average entanglement, $\langle Q\rangle$, over all 8-qubit 
initial computational basis states as a function of time for quantum sawtooth 
maps, $k = 1.5$ (chained line) and $k = -1.5$ (dotted line), and
Harper maps, $\gamma = 1$ (solid line) and $\gamma = .1$ (dashed line), 
compared to the random matrix average (horizontal dashed line). The chaotic 
maps quickly approach the random matrix average while the regular maps do not. 
The insets show matrix element distributions for the regular (light) and 
chaotic (dark) sawtooth maps at $t = 50$ (A), the regular 
(light) and chaotic (dark) Harper maps at $t = 50$ (B), and $t = 1$ (light) 
and 100 (dark) of the baker's map (C).
}
\end{figure}

Initial computational basis states evolved under the quantum baker's map 
attain $\langle Q\rangle$ values close to $\langle Q_{CUE}\rangle$ only at 
large $t$ \cite{Scott}. This is understood based on  the baker's map matrix 
element distrubtion which does not at all resemble $P_{CUE}(x)$, Fig. 
\ref{M4}C. However, for $t = 100$ the distribution is much closer to 
the CUE distribution. For an 8 qubit map $\langle Q(t = 1)\rangle$ 
is only .3080, compared to $\langle Q_{CUE} \rangle = .9883$, while 
$\langle Q(t = 100)\rangle$ is .9597. It is important 
to note that the quantum baker's map for Hilbert space dimensions which are 
a power of 2 is known to have an almost Poissonian nearest neighbor 
eigenvalue spectrum \cite{BV,sar}. Thus, at long times we see relatively high
entanglement generation without the presence of a Wigner-Dyson distribution. 
Rather, more iterations lead to increased matrix element randomness causing 
the greater entanglement generation.

We study two other examples of quantized chaotic maps: the quantum sawtooth 
map \cite{saw1,saw2},
\begin{equation}
U_{saw} = \frac{e^{-i\pi/4}}{\sqrt{N}}e^{ik\pi m^2/N}e^{i\pi(n-m)^2/N},
\end{equation}
and the quantum Harper map \cite{harper}, 
\begin{equation}
U_H = e^{iN\gamma \cos(2\pi q/N)}e^{iN\gamma \cos(2\pi p/N)}.
\end{equation}
All elements of the chaotic, $k = 1.5$, and regular, $k = -1.5$, sawtooth 
maps have equal amplitude. One iteration of either map on any computational 
basis state yields a state with $Q = 1$. For the chaotic sawtooth, the matrix 
element randomness increases with $t$, such that at $t = 50$ the matrix 
element distribution is practically $P_{CUE}(x)$ and 
$\langle Q(t = 50)\rangle = .98826$. For the regular sawtooth 
$\langle Q\rangle$ oscillates wildly as seen in figure \ref{M4}. This stems 
from the lack of an asymptotic randomness for the matrix elements.  

The matrix elements for the chaotic Harper, $\gamma = 1$, deviate only 
slightly from $P_{CUE}(x)$, and $\langle Q(t = 1)\rangle =.9814$. For 
$t = 50$  there is an increase in matrix element randomness and 
$\langle Q(t = 50)\rangle = .9882$. The regular Harper map, $\gamma = .1$, 
matrix element distribution and  $\langle Q\rangle$ also approach asymptotic 
limits as $t$ increases. These limits fall short of the random matrix
statistics but the average entanglement is still
$\langle Q(t\rightarrow\infty)\rangle \simeq .95$. Note that the asymptotic 
average entanglement of the regular Harper is about the same as that of the 
baker's map. In addition, the average entanglement after one iteration of the 
map is higher for the regular Harper than for the baker's map. This appears 
to be an exception to the conjecture that entanglement is a signature
of quantum chaos. The quantum baker's map is widely considered chaotic since 
it is the quantum analog of a chaotic map. The Harper's map for $k = -1.5$ is 
not chaotic since it is the quantum analog of a regular map. Yet the entangling
power of the Harper's map appears to be at least as good, if not better, than
that of the baker's map.

\begin{figure}
\includegraphics[height=5.8cm, width=8cm]{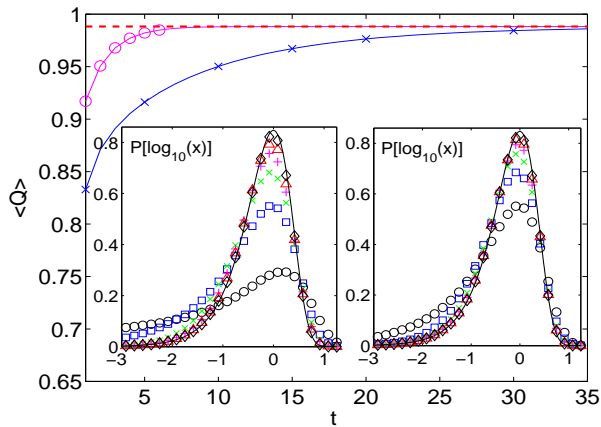}
\caption{\label{M3}
(Color online) Average entanglement generation as a function of time for 
50 interpolating ensemble operators with $\delta = .9$ ($\times$), and .98 
($\bigcirc$) operators on initial computational basis states. For both sets of 
operators the average entanglement produced approaches the CUE average (dashed
line) as an exponential. The insets show the matrix element distributions for 
the same set of operators as a function of time. The left inset shows the 
distribution for $\delta = .9$ operators for $t = 1 (\bigcirc)$, 5 ($\square$),
10 ($\times$), 15 (+), 20 ($\triangle$), and 30 ($\diamond$). The right inset 
shows the matrix element distributions for  $\delta = .98$ operators and 
$t = 1 (\bigcirc)$, 2 ($\square$), 3 ($\times$), 4 (+), 5 ($\triangle$), and 6 
($\diamond$). Both sets of operators matrix element distributions approach CUE 
as time increases. }
\end{figure}

Finally, we return to the interpolating ensembles matrix element distribution 
and $\langle Q\rangle$ now as a function of time. As shown in \cite{DK} we 
expect that the entanglement will approach the entanglement of the 
eigenvectors. However, we show here how matrix elements affect this and 
how operators with similar matrix element distributions lead to similar 
average entanglement generation. Fig. \ref{M3} shows $\langle Q(t)\rangle$ 
for $\delta = .9$ ($\times$), and .98 ($\bigcirc$) operators, already explored 
in \cite{YSW}, and the matrix element distribution at the same points in 
time. As the number of iterations increase the entanglement production 
exponentially approaches the CUE value (dashed line). The matrix element 
distribution also converges to the CUE distribution. Comparing the 
entanglement produced by these operators we note that 
$\langle Q(t = 5, 10, 15, 20, 30)\rangle $ of the $\delta = .9$ 
operators equal $\langle Q(t = 1, 2, 3, 4, 6)\rangle$, respectively, of the 
$\delta = .98$ operators. The matrix element distributions producing these 
entanglement values are practically equal. 

In conclusion, we have explored the connection between an operators matrix 
element distribution and its multi-partite entangling power on initial 
computational basis states. We have shown that operators with CUE distributions
of eigenvalues and eigenvectors are not the sole cause of CUE-like entangling 
power. CUE-like entangling power cannot be achieved without a CUE distribution 
of matrix elements. In addition, operators without CUE distributions of 
eigenvalues and eigenvectors can still have CUE-like entangling power if the 
operators have a random matrix element distribution. It also appears that 
operators with similar matrix element distributions generate similar amounts 
of entanglement. This analysis should provide a more specified goal in the search 
for efficient means of random state production.

The authors thank K. Zyczkowski for clarifying interpolating ensemble
generation and helpful discussions. The authors acknowledge support from
the DARPA QuIST (MIPR 02 N699-00) program. YSW acknowledges support
of the National Research Council through
the Naval Research Laboratory. Computations were performed at the ASC
DoD Major Shared Resource Center.

\appendix
\section{Appendix A}
CUE construction based on the Hurwitz parameterization starts with elementary 
unitary transformations, $E^{(i,j)}(\phi,\psi,\chi)$, with non-zero elements 
\cite{Zyc2}
\begin{eqnarray}
\label{E1}
E_{kk}^{(i,j)} &=& 1, \;\;\;\; k = 1, ... , N, \;\;\;\; k \neq i,j \nonumber\\
E_{ii}^{(i,j)} &=& e^{i\psi}\cos\phi, \;\;\;\;\;\;\;\;\; E_{ij}^{(i,j)} = e^{i\chi}\sin\phi \nonumber\\
E_{ji}^{(i,j)} &=& -e^{-i\chi}\sin\phi, \;\;\;\; E_{jj}^{(i,j)} = e^{-i\psi}\cos\phi
\end{eqnarray}
which are used to form $N-1$ composite rotations
\begin{eqnarray}
E_1 &=& E^{(N-1,N)}(\phi_{01},\psi_{01},\chi_1) \nonumber\\
E_2 &=& E^{(N-2,N-1)}(\phi_{12},\psi_{12},0)E^{(N-1,N)}(\phi_{02},\psi_{02},\chi_2) \nonumber\\
\dots\nonumber\\
E_{N-1} &=& E^{(1,2)}(\phi_{N-2,N-1},\psi_{N-2,N-1},0)\times\nonumber\\
 & & E^{(2,3)}(\phi_{N-3,N-1},\psi_{N-3,N-1},0)\times\nonumber\\
 &\dots& E^{(N-1,N)}(\phi_{0,N-1},\psi_{0,N-1},\chi_{N-1}).
\end{eqnarray}
A CUE matrix is finally attained by 
$U_{CUE} = e^{i\alpha}E_1E_2\dots E_{N-1}$. The Euler angles $\psi$, $\chi$, 
and $\alpha$ are drawn uniformly from the intervals
\begin{equation}
\label{E3}
0\leq \psi_{rs} \leq 2\pi, \;\;\;\;\;\; 0\leq \chi_{s} \leq 2\pi, \;\;\;\;\;\;
0\leq \alpha \leq 2\pi, 
\end{equation}
and $\phi_{rs} = \sin^{-1}({\xi_{rs}}^{1/(2r+2)})$, with $\xi_{rs}$ 
drawn uniformly from 0 to 1. The $2\times2$ block $E^{(i,j)}_{m,n}$ with 
$m,n = i,j$ and $r = 0$ is a random SU(2) rotation with respect to the Haar 
measure. The interpolating ensembles \cite{Zyc3} follow the same construction 
but the angles are drawn from constricted intervals
\begin{equation}
\label{delta}
0 \leq \psi_{rs} \leq 2\pi\delta, \;\;\;\;\;\; 0\leq \chi_s \leq 2\pi\delta, \;\;\;\;\;\; 0 \leq \alpha \leq 2\pi\delta, 
\end{equation}
with $\phi_{rs} = \sin^{-1}(\delta{\xi_{rs}}^{1/(2r+2)})$ and
$\xi_{rs}$ drawn from 0 to 1. The whole is multiplied by a diagonal matrix of 
random phases drawn uniformly from 0 to $2\pi$. The parameter $\delta$ 
ranges from 0 to 1 and provides a smooth transition of certain statistical 
properties between the diagonal circular Poisson ensemble (CPE) and CUE 
\cite{Zyc3}.

\end{document}